\documentclass[runningheads]{llncs}
\usepackage{booktabs}
\usepackage[T1]{fontenc}
\usepackage[utf8]{inputenc}
\usepackage{graphicx}
\usepackage{hyperref}
\usepackage{url}
\usepackage{todonotes}
\usepackage{wrapfig}
\usepackage{lipsum}
\usepackage{tikz}
\usetikzlibrary{arrows.meta, positioning, shapes.geometric, calc}
\usepackage{amsfonts}

\def\cC{{\mathcal{C}}}
\def\cD{{\mathcal{D}}}
\def\cF{{\mathcal{F}}}
\def\cG{{\mathcal{G}}}
\def\cU{{\mathcal{U}}}
\def\invto{{}}

\makeatletter
\renewcommand\section{\@startsection{section}{1}{\z@}%
                       {-12\p@ \@plus -4\p@ \@minus -4\p@}%
                       {8\p@ \@plus 4\p@ \@minus 4\p@}%
                       {\normalfont\large\bfseries\boldmath
                        \rightskip=\z@ \@plus 8em\pretolerance=10000 }}
\makeatother

\makeatletter
\renewcommand{\paragraph}[1]{%
  \@startsection{paragraph}{4}{0pt}{3pt}{0pt}{\itshape}%
  {#1}\hspace{1pt}
}
\makeatother

\begin{document}
\title{Exploring Formal Math on the Blockchain: An~Explorer for Proofgold}
\author{Chad E. Brown\inst{1} \and
  Cezary Kaliszyk\inst{2}\orcidID{0000-0002-8273-6059} \and
  Josef Urban\inst{1}\orcidID{0000-0002-1384-1613}
}
\authorrunning{C. Brown, C. Kaliszyk, J. Urban}
\institute{Czech Technical University in Prague, Czech Republic\\
\email{josef.urban@gmail.com} \and
  University of Melbourne, Australia and University of Innsbruck, Austria\\
\email{ckaliszyk@unimelb.edu.au}}
\maketitle
\begin{abstract}
Proofgold is a blockchain that supports formalized mathematics alongside standard cryptocurrency functionality. It incorporates logical constructs into the blockchain, including declarations of formal theories, definitions, propositions and proofs. It also supports placing and collecting bounties on proving these propositions, incentivizing the development of the formal libraries contained in Proofgold. In this paper, we present a web-based blockchain explorer for Proofgold. The system exposes not only the usual transactional data but also the formal mathematical components embedded in the chain and allows some interaction with them. The explorer allows users to inspect blocks, transactions, and addresses, as well as formal objects: theories, definitions, theorems and their proofs. We also support the submission of transactions to the blockchain using our interface.  We describe the system architecture and its integration with the Proofgold Lava software, highlighting how the explorer supports navigation of formal content and facilitates mathematical knowledge management in a decentralized setting, as well as a number of formalizations in category theory done in the system.
\end{abstract}


\section{Introduction}
Formalized mathematics has seen remarkable progress in recent years, with large-scale
developments such as the formal proof of the Feit-Thompson theorem in Coq~\cite{GonthierAABCGRMOBPRSTT13} and the proof of the Kepler conjecture in HOL Light and Isabelle~\cite{flyspeck-final},
the formalization of perfectoid spaces in Lean~\cite{BuzzardCM20},
and the formalization
of superposition calculus in Isabelle's Archive of Formal Proofs~\cite{DesharnaisTWBT24}. 
Despite these advances, most formalization efforts are still coordinated through centralized repositories,
and the ecosystem lacks well-established mechanisms for incentivizing contributions. Blockchain technology
offers a promising avenue to address both of these challenges by providing a decentralized infrastructure
for storing and verifying formal content, as well as built-in mechanisms for transparent incentives and
attribution. By combining blockchains with formalized mathematics, we can create systems in which formal
theories, definitions, propositions and their associated metadata are permanently recorded in a ledger.
This enables provenance tracking, ensures tamper resistance and facilitates collaborative contributions
across a global network without reliance on a central authority.

Proofgold~\cite{BrownKGU22} builds upon this idea of integrating formal mathematics with a decentralized system by providing
a blockchain that supports the storage, verification and incentivization of formal math contributions.
By including formal theories, definitions, theorems and proofs directly into the blockchain, Proofgold ensures
that this mathematical knowledge is publicly accessible and verifiable. By integrating a bounty system, where
users can place rewards for the verification (or refutation) of specific conjectures, it also
encourages development. Despite these advantages, Proofgold is currently a command-line interface-only system,
lacking a user interface, which limits its accessibility to experts who are familiar with the CLI and reduces
its appeal to a broader audience.

\paragraph{Contributions} In this paper, we present the Web-based blockchain explorer for Proofgold,
built by extending the Proofgold Lava client~\cite{BrownKGU22}. This extension adds various functionalities
for querying the blockchain, and includes the implementation of server-side code to facilitate interactions
with the client. In particular:

\begin{itemize}
\item We describe the system architecture, outlining how the web-based explorer interfaces with the Proofgold
  blockchain (Section~\ref{s:arch});
\item We demonstrate how formal mathematical objects -- such as theorems, definitions and proofs -- are rendered
  within the explorer, and explain how users can interact with these objects (Section~\ref{s:explorer});
\item As a case study, we present a number of conjectures in category theory along with their proofs or refutations
  presented in Proofgold (Section~\ref{s:usecase}).
\end{itemize}

\section{Proofgold and Formal Mathematics on the Blockchain}

This section briefly introduces Proofgold; for a more complete description, see \cite{BrownKGU22}.
Proofgold is a cryptocurrency designed to support formal logic and mathematics.
The core of Proofgold is a proof checker for intuitionistic higher-order logic with functional extensionality.
We only give an introduction to the higher-order Tarski Grothendieck set theory (HOTG) defined in Proofgold in Section~\ref{s:usecase}.
Users can publish theories, which consist of primitive constants, their types, and axioms. Each theory is uniquely identified by a
256-bit identifier derived from its recursive hash (Merkle root).
Documents defining new objects, proving theorems, and stating conjectures can be published within a theory. Ownership of propositions
is determined by public keys, enabling the redemption of bounties by proving conjectures. Proofgold combines proof-of-stake and
proof-of-burn, with the proof-of-burn element involving burning small amounts of Litecoin. This combination enhances security by
reusing Litecoin's proof-of-work. The first 5000 Proofgold blocks automatically placed bounties on pseudorandom propositions~\cite{BrownKGU22},
allowing new participants to increase their stake by proving theorems.

Proofgold has seen significant activity in terms of theories, documents, and formalizations. The platform includes a built-in
theory of hereditarily finite sets (HF), which was used to generate pseudorandom bounties for the first 5000 blocks. Additionally,
two theories axiomatizing HOTG have been published \cite{BrownP19}, one corresponding
to Mizar~\cite{BancerekBGKMNP18} and one corresponding to Megalodon, along with a theory
for reasoning
about syntax using higher-order abstract syntax (HOAS). These theories have facilitated the formalization of various mathematical
concepts and the construction of significant mathematical objects, such as the real numbers via Conway's surreal numbers.
In particular, the Megalodon proofs of 12 of Wiedijk's 100 theorems~\cite{Wiedijk06} are included in Proofgold.
Furthermore,
the platform has enabled the publication of conjectures and the collection of bounties, with 2836 of the 13142 bounties resolved
as of April 2025.

The Proofgold Lava Client, co-developed by the authors, addresses the scalability issues of the original Proofgold Core software.
It features an improved database layer using the Unix DBM interface, a more efficient cryptography layer, and enhancements to the
networking and proof-checking layers. The client is a command line interface that unfortunately requires somewhat complicated
installation (it relies on a litecoin running in RPC mode, and particular versions of the database). Additionally, it supports 210
commands, which may be somewhat overwhelming for users.

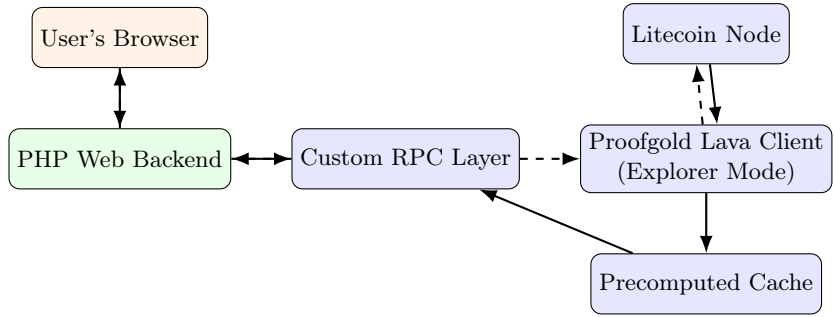
\begin{figure}[tb!]
\centering
\begin{tikzpicture}[
  node distance=.8cm and .8cm,
  box/.style={rectangle, draw, rounded corners, minimum height=.8cm, align=center, font=\small, fill=blue!10},
  backend/.style={rectangle, draw, rounded corners, minimum height=.8cm, align=center, font=\small, fill=green!10},
  frontend/.style={rectangle, draw, rounded corners, minimum height=.8cm, align=center, font=\small, fill=orange!10},
  arrow/.style={-Latex, thick},
]

\node[frontend] (frontend) {User's Browser};
\node[backend, below=of frontend] (backend) {PHP Web Backend};
\node[box, right=of backend] (rpc) {Custom RPC Layer};
\node[box, right=of rpc] (lava) {Proofgold Lava Client\\ (Explorer Mode)};
\node[box, below=of lava] (hashtables) {Precomputed Cache};
\node[box, above=of lava] (litecoin) {Litecoin Node};

\draw[arrow] ([xshift=15pt]litecoin) -- (lava);
\draw[arrow, dashed] ([xshift=-15pt]lava) -- (litecoin);
\draw[arrow] (lava) -- (hashtables);
\draw[arrow] (hashtables) -- (rpc);
\draw[arrow] (rpc) -- (backend);
\draw[arrow] (backend) -- (frontend);

\draw[arrow, dashed] (frontend) -- (backend);
\draw[arrow, dashed] (backend) -- (rpc);
\draw[arrow, dashed] (rpc) -- (lava);

\end{tikzpicture}
  \caption{\label{fig:arch}High-level architecture of the explorer}
\end{figure}

\section{System Architecture}\label{s:arch}

The design of the Proofgold Explorer focuses on providing accessible and interactive access to both blockchain
 data and formal mathematical content. In particular, we aim to offer: an intuitive web interface that lowers the entry barrier;
 structured access to formal and transactional data; and integration with the existing Proofgold infrastructure.
To achieve these goals, we propose the architecture consisting of three main components (Fig.~\ref{fig:arch}):
\begin{enumerate}
  \item A modified Proofgold Lava client, running in a new \texttt{explorer} mode that would cache and prepare additional information;
  \item A server-side backend written in PHP;
  \item An RPC communication layer connecting the backend with both the explorer node and the underlying Litecoin node.
\end{enumerate}

\begin{figure}[t!]
  \centerline{\includegraphics[width=\textwidth]{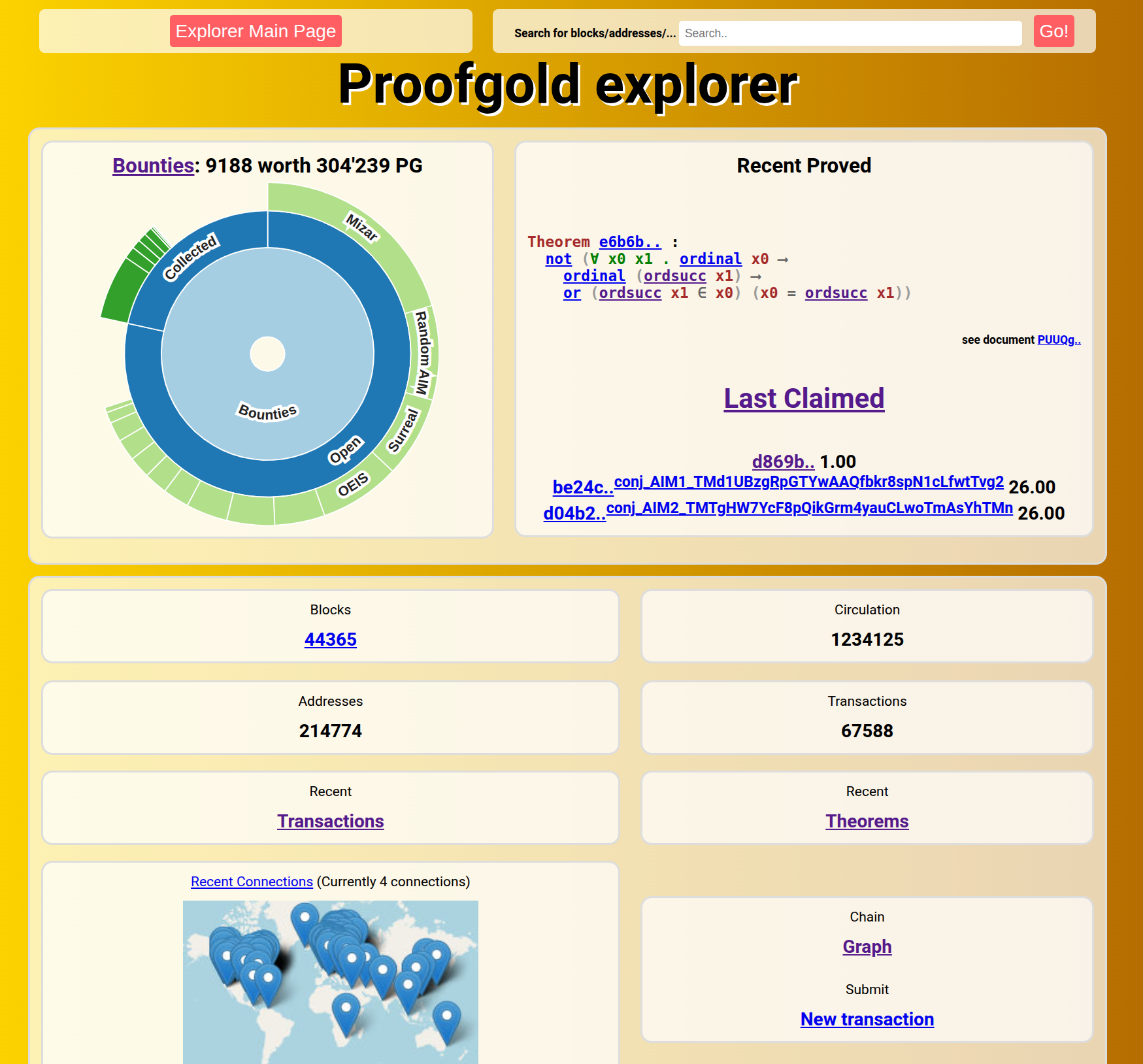}}
  \caption{\label{fig:main}Explorer Main Page}
\end{figure}

In explorer mode, the Lava client computes and caches additional information about the formal content embedded in the blockchain.
This includes the possibility of looking up definitions, propositions, theories, proofs and bounties. In principle, this information
is already stored in the blockchain, but it requires traversing the whole history. The added information is stored in 34 OCaml
hashtables that add approximately 10 Gigabytes in memory and that are periodically (every hour) refreshed.
These structures allow efficient access to: information about individual formal entities, including their types, statements, and owners;
the history and status of bounties and their collection;
dependencies between formal objects and their evolution over time;
ownership and authorship tracking for mathematical content.
 It serves this structured information through a custom RPC interface that returns it in a machine-readable format for use by the frontend.

The web backend is implemented in PHP and is responsible for serving user-facing pages that visualize both standard blockchain
 data (blocks, transactions, addresses, bounties) and formal mathematical constructs (theorems, definitions, proofs, theories, conjectures). The backend queries the explorer node in real time and formats the data into interactive HTML views. It also enables lightweight interaction with the blockchain, such as submitting transactions.
These components are designed to be kept in sync with the live blockchain state and support both incremental updates and history queries.

\begin{wrapfigure}{r}{0.45\textwidth}
  \vspace{-.5cm}
  \centerline{\includegraphics[height=14cm]{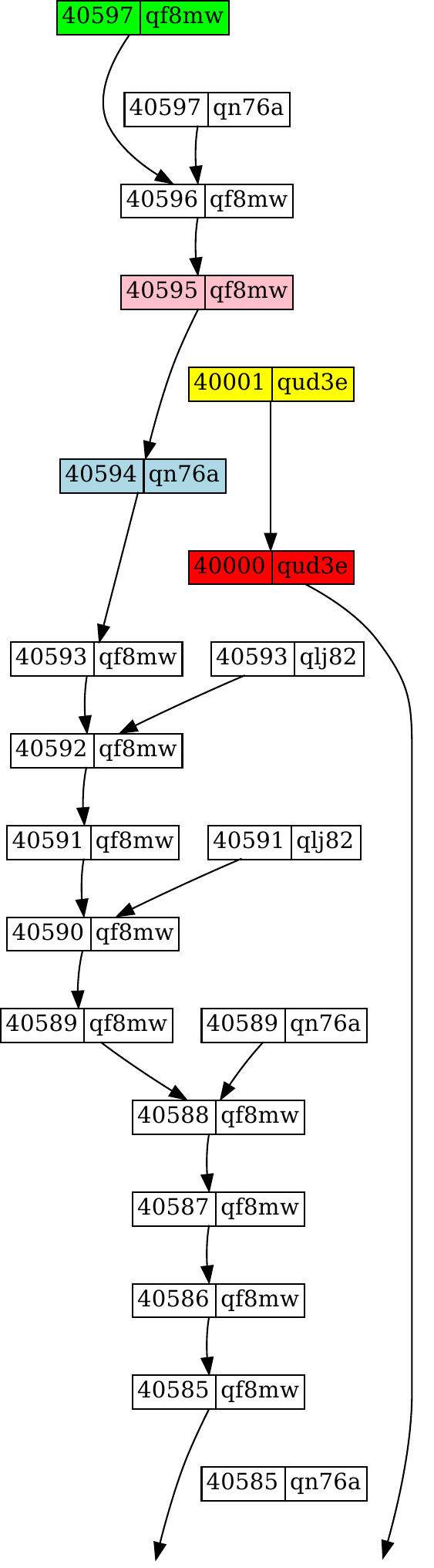}}
  \caption{\label{fig:graph}Chain Graph}
  \vspace{-.5cm}
\end{wrapfigure}

The explorer is also connected to the \textbf{Megalodon Wiki (mgwiki)}\footnote{\url{https://github.com/mgwiki/mgw_test}, \url{https://mgwiki.github.io/mgw_test/}} which is a collaborative git-based platform for formal math that enables users to edit and verify Megalodon  files directly in the browser. The mgwiki workflow involves cloning or forking the repository, modifying or adding \texttt{.mg} files, and committing changes, which triggers automated proof checking and HTML generation via GitHub Actions. Errors, if any, are reported in the GitHub action logs, and successful edits are published as browsable HTML. A unique feature of mgwiki is its integration with Proofgold: when a valid Megalodon file is added, it is automatically converted into a \texttt{.pfg} document that can be submitted to Proofgold. This allows users to associate formalized conjectures with bounties and track their progress through the explorer. Mgwiki thus serves also as a gateway for contributing conjectures and proofs to the Proofgold-based decentralized proof bounty system.

\section{Explorer Functionality}\label{s:explorer}

The Proofgold Explorer provides a structured and interactive interface to both blockchain
and formal mathematical data. The main dashboard (Figure~\ref{fig:main}), available online\footnote{\url{https://formalweb3.uibk.ac.at/pgbce/}, \url{http://proofgold.net/explorer/}} aggregates statistics
about the blockchain such as block height, address count, transaction volume and coin circulation.
An important feature is a link to the overview of the current graph of the blockchain. This is normally
a single chain with branches in the case of competing chain tips. We show an example from a time when
there were multiple competing nodes (Figure~\ref{fig:graph}). Particular nodes in the chain are
marked in special colors: nodes that define theories, include proof objects, place bounties
on propositions, or just include transactions are colored in green, blue, and pink. Additionally,
as the blockchain attempts to be resilient to attacks, the graph also marks missing nodes and
invalid nodes (spending non-existing assets, but also invalid proof steps) in yellow and red
respectively.

\begin{figure}[tb!]
  \centerline{\includegraphics[width=\textwidth]{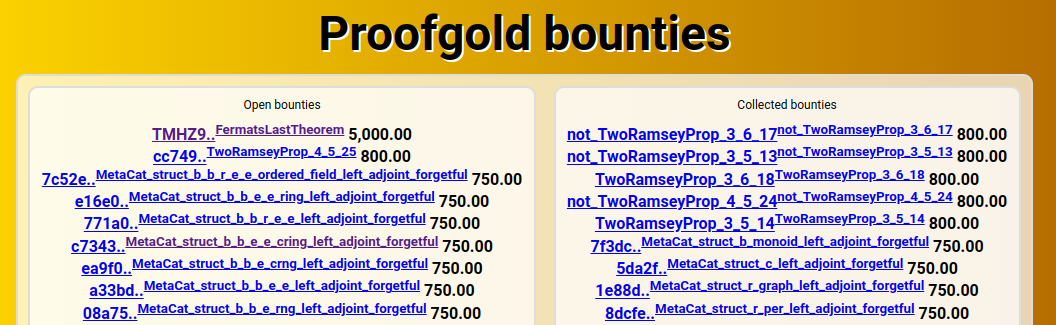}}
  \caption{\label{fig:bounties}Highest open and collected bounties}
\end{figure}

\begin{figure}[tb!]
  \centerline{\includegraphics[width=.62\textwidth]{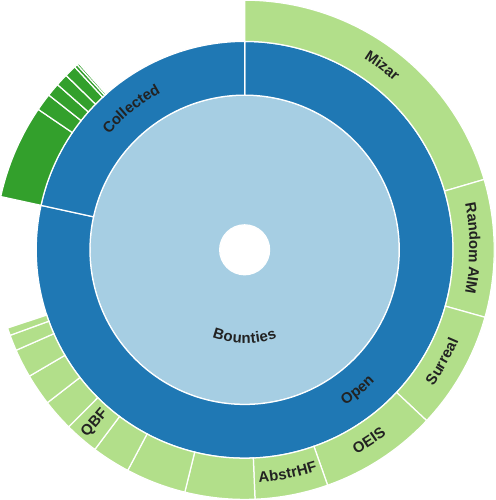}}
  \caption{\label{fig:bountycat}Open and collected bounty categories}
\end{figure}

\begin{figure}[tb!]
  \centerline{\includegraphics[width=\textwidth]{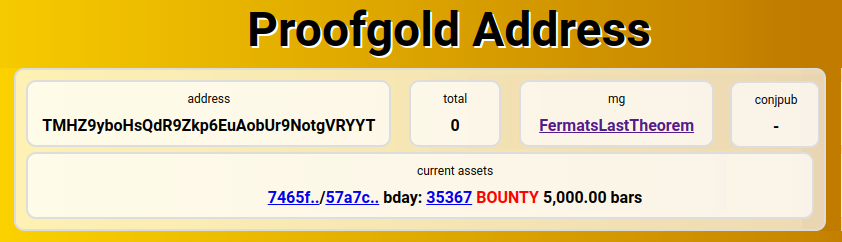}}
  \centerline{\includegraphics[width=\textwidth]{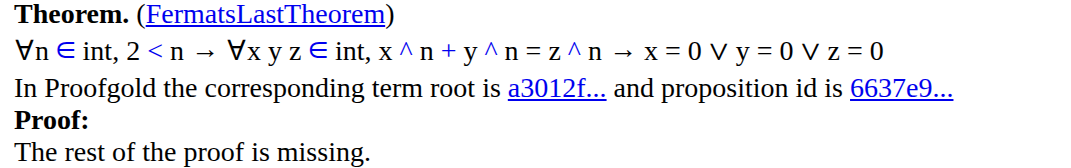}}
  \caption{\label{fig:fermat}Fermat as a bounty in the Proofgold explorer and its corresponding statement in the Megalodon Wiki}
\end{figure}

The explorer also visualizes the open and collected bounties, with a more detailed view of the
highest open and collected bounties, as well as a categorization of bounties presented in Figures
\ref{fig:bounties} and \ref{fig:bountycat}. This allows users to inspect the most valuable
open conjectures and focus their proof efforts on them, as well as to see what are the domains of
the conejctures users are working on.
The individual bounties can be viewed either in the Proofgold format, if their
representation is given in a theory document, or in case if they are only given in the
opaque formal, they are linked to the Megalodon Wiki.
We show this in the example
of Fermat's last theorem, as of April 2025 it is the conjecture with the highest bounty in Proofgold
(Figure~\ref{fig:fermat}).

\begin{figure}[tb!]
  \centerline{\includegraphics[width=\textwidth]{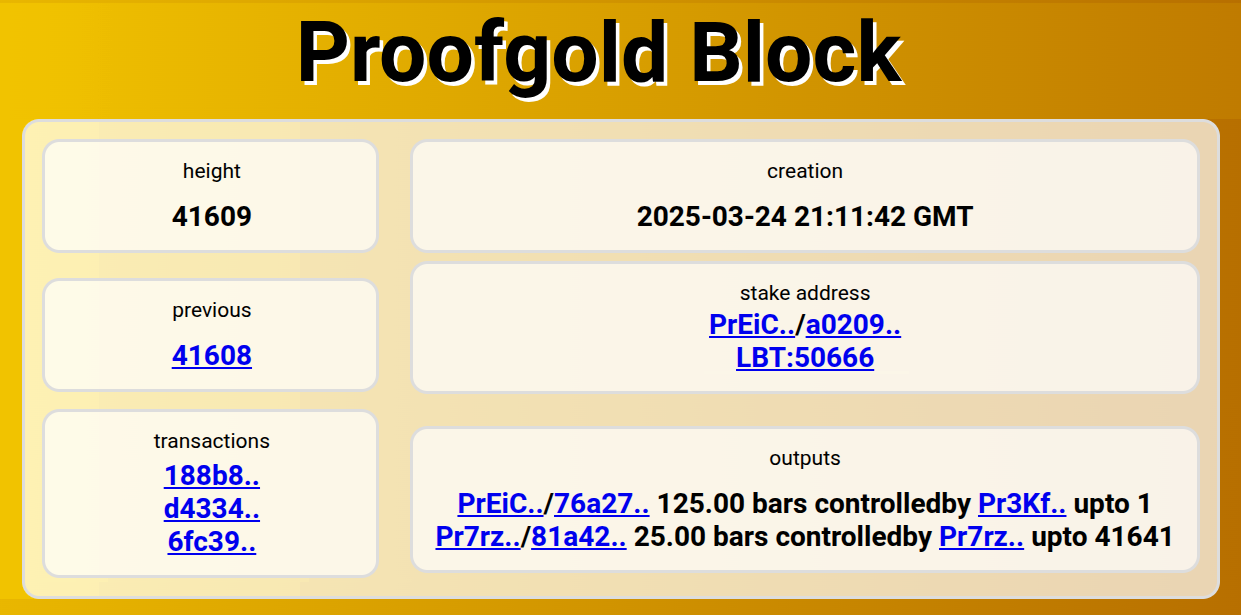}}
  \caption{\label{fig:block}Explorer view of a block}
\end{figure}

\begin{figure}[tb!]
  \centerline{\includegraphics[width=\textwidth]{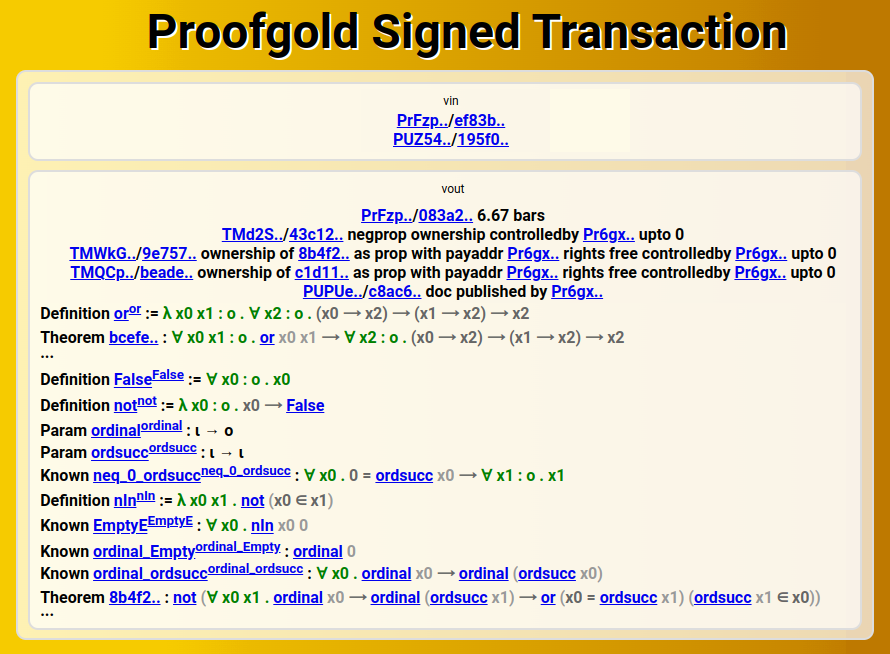}}
  \caption{\label{fig:stx}Explorer view of a transaction. Clicking ``\ldots'' allows inspecting the proof.}
\end{figure}

The explorer allows viewing blocks (Figure~\ref{fig:block}) and transactions in them. This is useful,
because unlike in other blockchains, Proofgold transactions are not always purely value-based. They
can also introduce proof documents. We show an example of
such a transaction in Figure~\ref{fig:stx}: The transaction takes a small amount of proofgold along
with a marker (used to safely claim ownership of proved objects \cite{BrownKGU22}) and sends
them to the address of a proof document. The newly defined objects, shown in the figure, and theorems
proved in this document are now owned by the publisher of the document.

\begin{figure}[tb!]
  \centerline{\includegraphics[width=\textwidth]{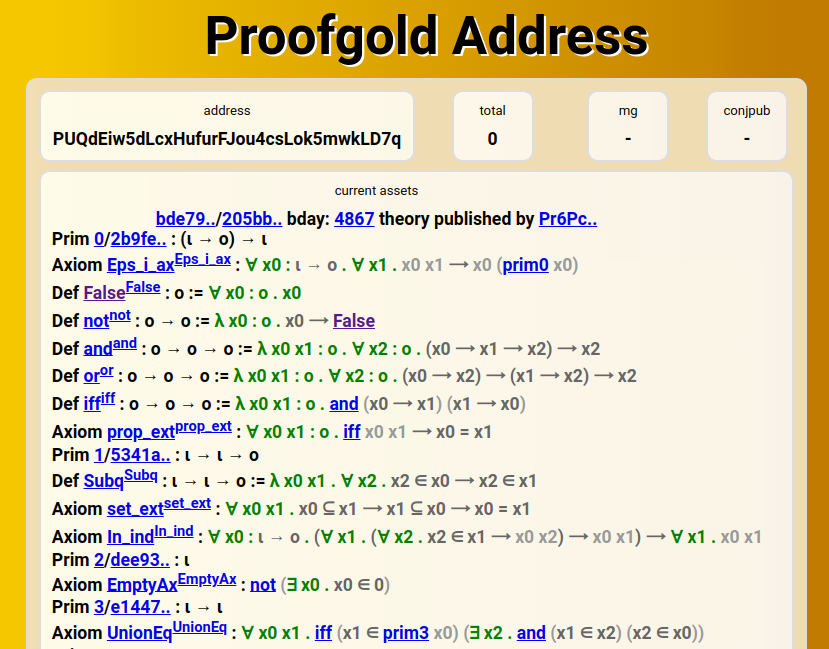}}
  \caption{\label{fig:hotg}The higher-order Tarski Groethendieck theory in the explorer}
\end{figure}

The explorer also allows viewing the list of all theories and viewing
individual ones. We show the theory defining document for the Tarski Groethendieck
theory in Fig.~\ref{fig:hotg}. The axioms directly corresponding to the foundations
of Megalodon as known from~\cite{BrownP19}.

\subsection{Transaction Submission Interface}

Blockchain explorers typically include transaction submission interfaces to allow users to directly
broadcast transactions to the network without needing a full node or wallet software. This is useful
for users who have already constructed and signed a transaction (usually offline) and simply need
to submit it to the blockchain for confirmation. We also include such an interface in the explorer.

\section{Use Case and Examples}\label{s:usecase}

Some of the examples of conjectures with bounties -- many of which have
already been proven or disproven -- assert the existence of
a left adjoint to a forgetful functor.
Such left adjoints, when they exist, often correspond to freely generated structures.
More generally, the interest in such adjoints can be justified by Slogan IV
of~\cite{LambScot86}: ``Many important concepts in mathematics arise as adjoints, right or left, to previously known functors.''
We begin with an informal description, leaving details to~\cite{LambScot86}.
Suppose $\cC$ and $\cD$ are categories
and $\cF:\cC\to\cD$ and $\cU:\cD\to\cC$ are functors.
$\cF$ is left adjoint to $\cU$ if there exist
natural transformations $\eta:1_\cC\to \cU\cF$ and $\varepsilon:\cF\cU\to 1_\cD$
satisfying certain identities.
In each conjecture below, $\cC$ will be the category of sets, $\cD$ will be some category of
structures and $\cU$ will be the forgetful functor sending a structure to its carrier set.
The conjecture will then state that there exist $\cF$, $\eta$ and $\varepsilon$
giving an adjunction.
In July 2021, 
bounties of 750 bars were placed on 33 conjectures of this form.
As of May 2025, 14 of the bounties have been collected (11 by proving the conjecture and 3 by
disproving the conjecture), and 19 remain open.
We discuss a few of these propositions below.

The conjectures and theorems are in the HOTG theory~\cite{BrownP19}, with a base type $\iota$ of sets.
We use $o$ for the type of propositions and $\alpha\beta$ for the type of functions from $\alpha$ to $\beta$.
We briefly review elements of set theory required to describe the conjectures.
There are the usual logical definitions: $\top:o$, $\bot:o$, $\lnot:o\invto o$, $\land:o\invto o\invto o$,
$\equiv:o\invto o \invto o$, $=$ and $\exists$.
Two primitives of the set theory are relevant: $\in:\iota\invto\iota\invto o$ (which we write in infix)
and $\emptyset:\iota$ (which we often write as $0$ below).
After enough infrastructure is defined, we also have the following objects:
\begin{itemize}
\item ${\mathsf{lam}}:\iota\invto (\iota\invto\iota)\invto\iota$: Here ${\mathsf{lam}}~X~f$ is the set encoding the function $f$ restricted to the domain $X$.
\item ${\mathsf{ap}}:\iota\invto\iota\invto\iota$: ${\mathsf{ap}}~f~x$ corresponds to applying the function (encoded by the set) $f$ to $x$. Here we often simply write $f~x$, leaving ${\mathsf{ap}}$ as implicit. (Note that since $f$ and $x$ have type $\iota$, $f~x$ would be ill-typed if we did not insert ${\mathsf{ap}}$.)
\item ${\mathsf{Pi}}:\iota\invto (\iota\invto\iota)\invto\iota$: Here $\Pi~X~Y$ is the set of functions $f:X\to \bigcup_{x\in X}(Y~x)$ such that $f~x\in Y~x$ for each $x\in X$. We write $Y^X$ for the term $\Pi~X~(\lambda x.Y)$. Note that $Y^X$ is simply the set of functions from $X$ to $Y$.
\end{itemize}
For the categories of interest, the following definitions are important.
\begin{itemize}
\item ${\mathsf{lam\_id}}:\iota\invto\iota$ where ${\mathsf{lam\_id}}~X$ is ${\mathsf{lam}}~X~(\lambda x.x)$. That is, ${\mathsf{lam\_id}}$ is the set-theoretic encoding of the identity function on $X$.
\item ${\mathsf{lam\_comp}}:\iota\invto\iota\invto\iota\invto\iota$ where ${\mathsf{lam\_comp}}~X~g~f$ is ${\mathsf{lam}}~X~(\lambda x.g (f x))$. Assuming $f$ is a function with domain $X$ and $g$ is appropriate, this is the set-theoretic encoding of the composition of $f$ and $g$.
\item ${\mathsf{HomSet}}:\iota\invto\iota\invto\iota\invto o$ where ${\mathsf{HomSet}}~X~Y~f$ is $f\in Y^X$.
\end{itemize}
Some previously proven results can be assumed here, e.g.,
$\forall X.{\mathsf{lam\_id}}~X\in X^X$
and
$\forall X Y f.f\in Y^X\to {\mathsf{lam\_comp}}~X~f~({\mathsf{lam\_id}}~X) = f$.

We will only consider structures with a single carrier set
and assume the set theoretic representation of each structure is as a function $A$ where $A~0$ (i.e., $A$ applied to $0$)
yields the carrier.
To account for this, the identities and composition for categories of structures
are defined by slight modifications of ${\mathsf{lam\_id}}$ and ${\mathsf{lam\_comp}}$.
\begin{itemize}
\item ${\mathsf{struct\_id}}:\iota\invto\iota$ where ${\mathsf{struct\_id}}~A$ is ${\mathsf{lam\_id}}~(A~0)$.
\item ${\mathsf{struct\_comp}}:\iota\invto\iota\invto\iota\invto\iota\invto\iota\invto\iota$ where
  ${\mathsf{struct\_comp}}~A~B~C$ is ${\mathsf{lam\_comp}}~(A~0)$.
  Note that this ignores its second and third arguments. Giving the last two arguments explicitly yields
  ${\mathsf{struct\_comp}}~A~B~C~g~f$ is ${\mathsf{lam\_comp}}~(A~0)~g~f$.
\end{itemize}

There are many examples of structures defined in Megalodon and Proofgold
which differ in two ways: the other components of the structure (a binary operation, a binary relation, etc.)
and what properties are assumed of these other components.
To give a simple, concrete example, we consider structures with a single binary relation.
Avoiding details (which are not relevant here), we note that there is a previously defined
object ${\mathsf{pack\_r}}:\iota\invto (\iota\invto\iota\invto o)\invto \iota$
such that ${\mathsf{pack\_r}}~X~R$ encodes the carrier set $X$ and the binary relation $R$
(restricted to its behavior on $X$) as a set.
We then define ${\mathsf{struct\_r}}:\iota\invto o$ to be the class of all such sets.
We also assume the previously
proven $\forall X.\forall R:\iota\invto\iota\invto o. X = {\mathsf{pack\_r}}~X~R~0$.
That is, the set ${\mathsf{pack\_r}}~X~R$ is a function that yields the carrier set $X$ when applied to $0$.

A particular category of such structures will result from adding some restriction of the class ${\mathsf{struct\_r}}$
as objects.
Regardless of the restriction, the arrows should be all functions sending related inputs to related outputs.
This is given by ${\mathsf{BinRelnHom}}:\iota\invto\iota\invto\iota\invto o$
and the previously proven identity
$$
\begin{array}{c}
  {\mathsf{BinRelnHom}}~({\mathsf{pack\_r}}~X~R)~({\mathsf{pack\_r}}~Y~Q)~h \\
  = (h \in Y^X \land \forall x y \in X.R~x~y\to Q~(h~x)~(h~y)).
\end{array}$$
To have a specific category as an example, we consider ${\mathsf{IrrPartOrd}}:\iota\invto o$
which is the class of all structures with a single binary irreflexive transitive relation (i.e., an irreflexive partial order).
The details of the definition are not important. It is enough to know we have the following
previously proven results:
\begin{itemize}
\item If $R$ is irreflexive and transitive on $X$, then ${\mathsf{IrrPartOrd}}~({\mathsf{pack\_r}}~X~R)$.
\item Let $A$ satisfy ${\mathsf{IrrPartOrd}}~A$ and $q:\iota\invto o$ be given.
  In order to prove $q~A$ it is enough to prove $q~({\mathsf{pack\_r}}~X~R)$ for all sets $X$
  and irreflexive transitive relations $R$ on $X$.
\end{itemize}

Let us now turn to the relevant formalization of category theory.
In practice, we will be interested in large categories (metalevel categories),
so formally $\cC$ will not be encoded as a set
but will consist of four components (given as four explicit dependencies
in the formalization):
\begin{itemize}
\item $\cC_0:\iota\invto o$ -- the class of all objects of $\cC$.
\item $\cC_1:\iota\invto\iota\invto\iota\invto o$ -- where $\cC_1~X~Y$ is the class of arrows from $X$ to $Y$ in $\cC$.
\item ${\mathsf{id}}_\cC:\iota\invto\iota$ -- where ${\mathsf{id}}_\cC~X$ is the identity arrow for $X$ in $\cC$.
\item ${\mathsf{comp}}_\cC:\iota\invto\iota\invto\iota\invto\iota\invto\iota\invto\iota$ -- where ${\mathsf{comp}}_\cC~X~Y~Z~g~f$ is the composition of $f$ (an arrow from $X$ to $Y$) and $g$ (an arrow from $Y$ to $Z$).
\end{itemize}
The second category $\cD$ will also be represented by four explicit components: $\cD_0$, $\cD_1$, ${\mathsf{id}}_\cD$ and ${\mathsf{comp}}_\cD$.
Similarly, the functor $\cF$ is represented by two explicit components:
\begin{itemize}
\item $\cF_0:\iota\invto\iota$ -- where $\cF_0~X$ is the object of $\cD$ to which the object $X$ of $\cC$ maps.
\item $\cF_1:\iota\invto\iota\invto\iota\invto\iota$ -- where $\cF_1~X~Y~f$ is the arrow from $\cF_0~X$ to $\cF_1~Y$ in $\cD$ corresponding to the arrow $f$ from $X$ to $Y$ in $\cC$.
\end{itemize}
Of course, the functor $\cU$ is also represented by $\cU_0$ and $\cU_1$.
A natural transformation $\eta$ is simply of type $\iota\invto\iota$ -- mapping an object in one
category to an appropriate arrow in the other category.

${\mathsf{MetaCat}}:(\iota\invto o)\invto (\iota\invto\iota\invto\iota\invto o)\invto (\iota\invto\iota)\invto (\iota\invto\iota\invto\iota\invto\iota\invto\iota\invto\iota)\invto o$ is defined so that
${\mathsf{MetaCat}}~\cC_0~\cC_1~{\mathsf{id}}_\cC~{\mathsf{comp}}_\cC$
holds if the components form a category. Here, we only need to know that two specific (alleged) categories are categories.
The category of sets is given by the constant true predicate (every set is an object),
${\mathsf{HomSet}}$, ${\mathsf{lam\_id}}$ and $(\lambda X Y Z.{\mathsf{lam\_comp}}~X)$
(where the ignored arguments are needed for the types to match).
This has been previously proven to be a category, and the precise proposition is
${\mathsf{MetaCat}}~(\lambda X.\top)~{\mathsf{HomSet}}~{\mathsf{lam\_id}}~(\lambda X Y Z.{\mathsf{lam\_comp}}~X)$.
Likewise, the category of all irreflexive transitive relations has been proven to be a category
and the proposition is
${\mathsf{MetaCat}}~{\mathsf{IrrPartOrd}}~{\mathsf{BinRelnHom}}~{\mathsf{struct\_id}}~{\mathsf{struct\_comp}}$.

We know $\cF_0$ and $\cF_1$ give a functor between two categories if a number of basic properties hold.
Without going into details, there is a previous definition ${\mathsf{MetaFunctor}}$
such that ${\mathsf{MetaFunctor}}~\cC_0~\cC_1~{\mathsf{id}}_\cC~{\mathsf{comp}}_\cC~\cD_0~\cD_1~{\mathsf{id}}_\cD~{\mathsf{comp}}_\cD~\cF_0~\cF_1$ holds
precisely if those properties hold.
A previously proven result allows us to infer
${\mathsf{MetaFunctor}}~\cC_0~\cC_1~{\mathsf{id}}_\cC~{\mathsf{comp}}_\cC~\cD_0~\cD_1~{\mathsf{id}}_\cD~{\mathsf{comp}}_\cD~\cF_0~\cF_1$
by proving those properties.
There is also an object ${\mathsf{MetaFunctor\_strict}}$ which further requires the
two (alleged) categories to actually be categories.

We already know the forgetful functor from the category of irreflexive transitive relations
to the category of sets is a functor.
The forgetful functor (in general) sends a structure $A$ to its carrier set $A~0$
and sends structure morphisms $f$ to $f$ (which is already an appropriate set-theoretic function).
Hence, the previously proven theorem is
$$
\begin{array}{c}
  {\mathsf{MetaFunctor}}~{\mathsf{IrrPartOrd}}~{\mathsf{BinRelnHom}}~{\mathsf{struct\_id}}~{\mathsf{struct\_comp}}\\
  (\lambda X.\top)~{\mathsf{HomSet}}~{\mathsf{lam\_id}}~(\lambda X Y Z.{\mathsf{lam\_comp}}~X)\\
  (\lambda A.A~0)~(\lambda A B f.f).
\end{array}
  $$

Similar to the discussion above, we have ${\mathsf{MetaNatTrans}}$ where
$${\mathsf{MetaNatTrans}}~\cC_0~\cC_1~{\mathsf{id}}_\cC~{\mathsf{comp}}_\cC~\cD_0~\cD_1~{\mathsf{id}}_\cD~{\mathsf{comp}}_\cD~\cF_0~\cF_1~\cG_0~\cG_1~\eta$$
holds if $\eta$ is a natural transformation from $\cF$ to $\cG$ (assuming $\cC$, $\cD$, $\cF$ and $\cG$ are appropriate inputs).
Finally we have ${\mathsf{MetaAdjunction}}$ where
$${\mathsf{MetaAdjunction}}~\cC_0~\cC_1~{\mathsf{id}}_\cC~{\mathsf{comp}}_\cC~\cD_0~\cD_1~{\mathsf{id}}_\cD~{\mathsf{comp}}_\cD~\cF_0~\cF_1~\cU_0~\cU_1~\eta~\varepsilon$$
is an adjunction (with witnessing natural transformations $\eta$ and $\varepsilon$)
-- again, assuming the inputs are appropriate.
We also have ${\mathsf{MetaAdjunction\_strict}}$ which further ensures
the inputs are appropriate by requiring:
\begin{itemize}
\item ${\mathsf{MetaFunctor\_strict}}~\cC_0~\cC_1~{\mathsf{id}}_\cC~{\mathsf{comp}}_\cC~\cD_0~\cD_1~{\mathsf{id}}_\cD~{\mathsf{comp}}_\cD~\cF_0~\cF_1$ -- ensuring $\cC$ and $\cD$ are categories and $\cF$ is a functor from $\cC$ to $\cD$.
\item ${\mathsf{MetaFunctor}}~\cD_0~\cD_1~{\mathsf{id}}_\cD~{\mathsf{comp}}_\cD~\cC_0~\cC_1~{\mathsf{id}}_\cC~{\mathsf{comp}}_\cC~\cU_0~\cU_1$ -- ensuring $\cU$ is a functor from $\cD$ to $\cC$.
\item 
   ${\mathsf{MetaNatTrans}}~\cdots~\eta$
  -- ensuring $\eta$ is a natural transformation from $1_\cC$ to $\cU\cF$.
\item 
    ${\mathsf{MetaNatTrans}}~\cdots~\varepsilon$
  -- ensuring $\varepsilon$ is a natural transformation from $\cF\cU$ to $1_\cD$.
\end{itemize}

Now, for each category $\cD$ of structures, we can form the following conjecture that
a left adjoint to the forgetful functor exists:
$$
\begin{array}{c}
\exists \cF_0:\iota\invto\iota. \exists \cF_1:\iota\invto\iota\invto\iota\invto\iota.\exists \eta \varepsilon:\iota\invto\iota.\\
        {\mathsf{MetaAdjunction\_strict}}~(\lambda X.\top)~{\mathsf{HomSet}}~{\mathsf{lam\_id}}~(\lambda X Y Z.{\mathsf{lam\_comp}}~X)\\
        \cD_0~\cD_1~{\mathsf{struct\_id}}~{\mathsf{struct\_comp}}\\
        \cF_0~\cF_1~(\lambda A.A~0)~(\lambda A B f.f)~\eta~\varepsilon.
\end{array}
$$
As mentioned above, bounties were placed on 33 propositions of this form,
of which 19 remain unresolved (neither proven nor disproven).
Table~\ref{tab:adjforget} lists the 33 propositions and indications whether
the proposition has been proven, disproven or is still open.
In the particular case of irreflexive transitive relations,
$\cD_0$ is ${\mathsf{IrrPartOrd}}$
and $\cD_1$ is ${\mathsf{BinRelnHom}}$.
The proof of this special case is given by taking $\cF_0$ to be
$\lambda X.{\mathsf{pack\_r}}~X~(\lambda x y.\bot)$ -- that is, a set $X$
is taken to the structure given by $X$ with the empty relation.
Furthermore $\cF_1$ is given so that $\cF_1~X~Y~f$ is $f$,
$\eta$ is given so that $\eta~X$ is ${\mathsf{lam\_id}}~X$
and
$\varepsilon$ is given so that $\varepsilon~A$ is ${\mathsf{lam\_id}}~(A~0)$.
The remainder of the proof involves checking the relevant properties.
This proof was published on the Proofgold blockchain in September 2021.

\begin{table}
  \begin{center}
  \begin{tabular}[]{c|c}
    Name & Status \\ \hline
    \href{https://mgwiki.github.io/mgw\_test/conj/cat/Category\_struct\_p.mg.html#MetaCat\_struct\_p\_left\_adjoint\_forgetful}{MetaCat\_struct\_p\_left\_adjoint\_forgetful}
    & Proven \\
    \href{https://mgwiki.github.io/mgw\_test/conj/cat/Category\_struct\_r.mg.html#MetaCat\_struct\_r\_left\_adjoint\_forgetful}{MetaCat\_struct\_r\_left\_adjoint\_forgetful}
    & Proven \\
    \href{https://mgwiki.github.io/mgw\_test/conj/cat/Category\_struct\_r\_graph.mg.html#MetaCat\_struct\_r\_graph\_left\_adjoint\_forgetful}{MetaCat\_struct\_r\_graph\_left\_adjoint\_forgetful}
    & Proven \\
    \href{https://mgwiki.github.io/mgw\_test/conj/cat/Category\_struct\_r\_partialord.mg.html#MetaCat\_struct\_r\_partialord\_left\_adjoint\_forgetful}{MetaCat\_struct\_r\_partialord\_left\_adjoint\_forgetful}
    & Proven \\
    \href{https://mgwiki.github.io/mgw\_test/conj/cat/Category\_struct\_r\_ord.mg.html#MetaCat\_struct\_r\_ord\_left\_adjoint\_forgetful}{MetaCat\_struct\_r\_ord\_left\_adjoint\_forgetful}
    & Disproven \\
    \href{https://mgwiki.github.io/mgw\_test/conj/cat/Category\_struct\_r\_wellord.mg.html#MetaCat\_struct\_r\_wellord\_left\_adjoint\_forgetful}{MetaCat\_struct\_r\_wellord\_left\_adjoint\_forgetful}
    & Disproven \\
    \href{https://mgwiki.github.io/mgw\_test/conj/cat/Category\_struct\_r\_per.mg.html#MetaCat\_struct\_r\_per\_left\_adjoint\_forgetful}{MetaCat\_struct\_r\_per\_left\_adjoint\_forgetful}
    & Proven \\
    \href{https://mgwiki.github.io/mgw\_test/conj/cat/Category\_struct\_r\_equivreln.mg.html#MetaCat\_struct\_r\_equivreln\_left\_adjoint\_forgetful}{MetaCat\_struct\_r\_equivreln\_left\_adjoint\_forgetful}
    & Proven \\
    \href{https://mgwiki.github.io/mgw\_test/conj/cat/Category\_struct\_c.mg.html#MetaCat\_struct\_c\_left\_adjoint\_forgetful}{MetaCat\_struct\_c\_left\_adjoint\_forgetful}
    & Proven \\
    \href{https://mgwiki.github.io/mgw\_test/conj/cat/Category\_struct\_c\_topology.mg.html#MetaCat\_struct\_c\_topology\_left\_adjoint\_forgetful}{MetaCat\_struct\_c\_topology\_left\_adjoint\_forgetful}
    & Open \\
    \href{https://mgwiki.github.io/mgw\_test/conj/cat/Category\_struct\_c\_T1\_topology.mg.html#MetaCat\_struct\_c\_T1\_topology\_left\_adjoint\_forgetful}{MetaCat\_struct\_c\_T1\_topology\_left\_adjoint\_forgetful}
    & Open \\
    \href{https://mgwiki.github.io/mgw\_test/conj/cat/Category\_struct\_c\_Hausdorff\_topology.mg.html#MetaCat\_struct\_c\_Hausdorff\_topology\_left\_adjoint\_forgetful}{MetaCat\_struct\_c\_Hausdorff\_topology\_left\_adjoint\_forgetful}
    & Open \\
    \href{https://mgwiki.github.io/mgw\_test/conj/cat/Category\_struct\_u.mg.html#MetaCat\_struct\_u\_left\_adjoint\_forgetful}{MetaCat\_struct\_u\_left\_adjoint\_forgetful}
    & Proven \\
    \href{https://mgwiki.github.io/mgw\_test/conj/cat/Category\_struct\_u\_inj.mg.html#MetaCat\_struct\_u\_inj\_left\_adjoint\_forgetful}{MetaCat\_struct\_u\_inj\_left\_adjoint\_forgetful}
    & Proven \\
    \href{https://mgwiki.github.io/mgw\_test/conj/cat/Category\_struct\_u\_bij.mg.html#MetaCat\_struct\_u\_bij\_left\_adjoint\_forgetful}{MetaCat\_struct\_u\_bij\_left\_adjoint\_forgetful}
    & Proven \\
    \href{https://mgwiki.github.io/mgw\_test/conj/cat/Category\_struct\_u\_idem.mg.html#MetaCat\_struct\_u\_idem\_left\_adjoint\_forgetful}{MetaCat\_struct\_u\_idem\_left\_adjoint\_forgetful}
    & Proven \\
    \href{https://mgwiki.github.io/mgw\_test/conj/cat/Category\_struct\_b.mg.html#MetaCat\_struct\_b\_left\_adjoint\_forgetful}{MetaCat\_struct\_b\_left\_adjoint\_forgetful}
    & Open \\
    \href{https://mgwiki.github.io/mgw\_test/conj/cat/Category\_struct\_b\_quasigroup.mg.html#MetaCat\_struct\_b\_quasigroup\_left\_adjoint\_forgetful}{MetaCat\_struct\_b\_quasigroup\_left\_adjoint\_forgetful}
    & Open \\
    \href{https://mgwiki.github.io/mgw\_test/conj/cat/Category\_struct\_b\_loop.mg.html#MetaCat\_struct\_b\_loop\_left\_adjoint\_forgetful}{MetaCat\_struct\_b\_loop\_left\_adjoint\_forgetful}
    & Open \\
    \href{https://mgwiki.github.io/mgw\_test/conj/cat/Category\_struct\_b\_semigroup.mg.html#MetaCat\_struct\_b\_semigroup\_left\_adjoint\_forgetful}{MetaCat\_struct\_b\_semigroup\_left\_adjoint\_forgetful}
    & Open \\
    \href{https://mgwiki.github.io/mgw\_test/conj/cat/Category\_struct\_b\_monoid.mg.html#MetaCat\_struct\_b\_monoid\_left\_adjoint\_forgetful}{MetaCat\_struct\_b\_monoid\_left\_adjoint\_forgetful}
    & Disproven \\
    \href{https://mgwiki.github.io/mgw\_test/conj/cat/Category\_struct\_b\_group.mg.html#MetaCat\_struct\_b\_group\_left\_adjoint\_forgetful}{MetaCat\_struct\_b\_group\_left\_adjoint\_forgetful}
    & Open \\
    \href{https://mgwiki.github.io/mgw\_test/conj/cat/Category\_struct\_b\_abelian\_group.mg.html#MetaCat\_struct\_b\_abelian\_group\_left\_adjoint\_forgetful}{MetaCat\_struct\_b\_abelian\_group\_left\_adjoint\_forgetful}
    & Open \\
    \href{https://mgwiki.github.io/mgw\_test/conj/cat/Category\_struct\_b\_b\_e.mg.html#MetaCat\_struct\_b\_b\_e\_left\_adjoint\_forgetful}{MetaCat\_struct\_b\_b\_e\_left\_adjoint\_forgetful}
    & Open \\
    \href{https://mgwiki.github.io/mgw\_test/conj/cat/Category\_struct\_b\_b\_e\_rng.mg.html#MetaCat\_struct\_b\_b\_e\_rng\_left\_adjoint\_forgetful}{MetaCat\_struct\_b\_b\_e\_rng\_left\_adjoint\_forgetful}
    & Open \\
    \href{https://mgwiki.github.io/mgw\_test/conj/cat/Category\_struct\_b\_b\_e\_crng.mg.html#MetaCat\_struct\_b\_b\_e\_crng\_left\_adjoint\_forgetful}{MetaCat\_struct\_b\_b\_e\_crng\_left\_adjoint\_forgetful}
    & Open \\
    \href{https://mgwiki.github.io/mgw\_test/conj/cat/Category\_struct\_b\_b\_e\_e.mg.html#MetaCat\_struct\_b\_b\_e\_e\_left\_adjoint\_forgetful}{MetaCat\_struct\_b\_b\_e\_e\_left\_adjoint\_forgetful}
    & Open \\
    \href{https://mgwiki.github.io/mgw\_test/conj/cat/Category\_struct\_b\_b\_e\_e\_semiring.mg.html#MetaCat\_struct\_b\_b\_e\_e\_semiring\_left\_adjoint\_forgetful}{MetaCat\_struct\_b\_b\_e\_e\_semiring\_left\_adjoint\_forgetful}
    & Open \\
    \href{https://mgwiki.github.io/mgw\_test/conj/cat/Category\_struct\_b\_b\_e\_e\_ring.mg.html#MetaCat\_struct\_b\_b\_e\_e\_ring\_left\_adjoint\_forgetful}{MetaCat\_struct\_b\_b\_e\_e\_ring\_left\_adjoint\_forgetful}
    & Open \\
    \href{https://mgwiki.github.io/mgw\_test/conj/cat/Category\_struct\_b\_b\_e\_e\_cring.mg.html#MetaCat\_struct\_b\_b\_e\_e\_cring\_left\_adjoint\_forgetful}{MetaCat\_struct\_b\_b\_e\_e\_cring\_left\_adjoint\_forgetful}
    & Open \\
    \href{https://mgwiki.github.io/mgw\_test/conj/cat/Category\_struct\_b\_b\_e\_e\_field.mg.html#MetaCat\_struct\_b\_b\_e\_e\_field\_left\_adjoint\_forgetful}{MetaCat\_struct\_b\_b\_e\_e\_field\_left\_adjoint\_forgetful}
    & Open \\
    \href{https://mgwiki.github.io/mgw\_test/conj/cat/Category\_struct\_b\_b\_r\_e\_e.mg.html#MetaCat\_struct\_b\_b\_r\_e\_e\_left\_adjoint\_forgetful}{MetaCat\_struct\_b\_b\_r\_e\_e\_left\_adjoint\_forgetful}
    & Open \\
    \href{https://mgwiki.github.io/mgw\_test/conj/cat/Category\_struct\_b\_b\_r\_e\_e\_ordered\_field.mg.html#MetaCat\_struct\_b\_b\_r\_e\_e\_ordered\_field\_left\_adjoint\_forgetful}{MetaCat\_struct\_b\_b\_r\_e\_e\_ordered\_field\_left\_adjoint\_forgetful}
    & Open \\
  \end{tabular}
  \end{center}
  \caption{\label{tab:adjforget} Propositions asserting existence of adjunctions for forgetful functors}
\end{table}

\begin{figure}[tb!]
  \centerline{\includegraphics[width=\textwidth]{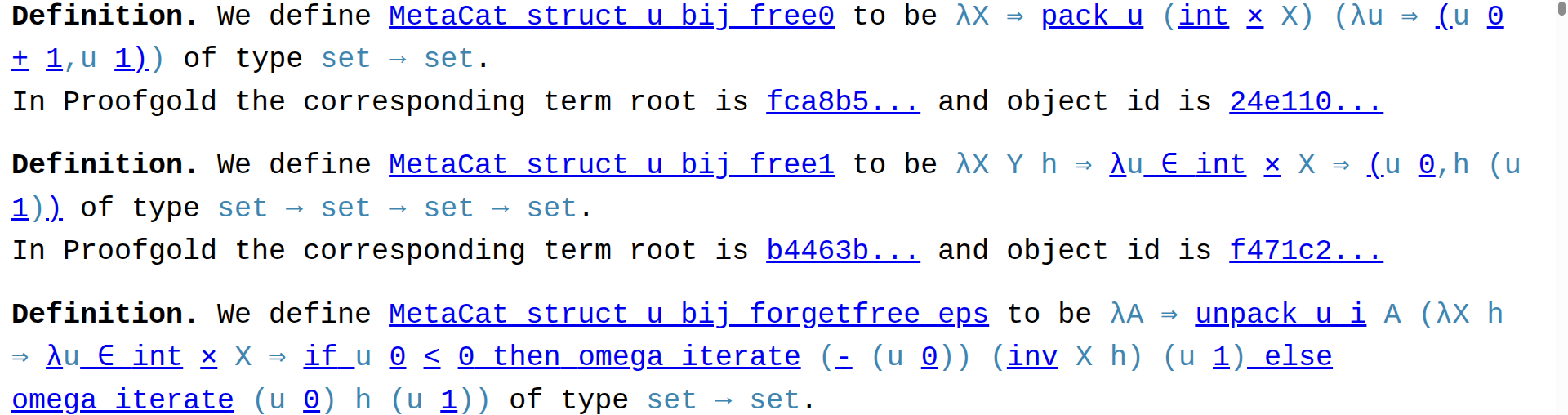}}
  \caption{\label{fig:bijfreedefs} Definitions for the Adjoint Functor for Bijections in the Megalodon Wiki}
\end{figure}

\begin{figure}[tb!]
  \centerline{\includegraphics[width=\textwidth]{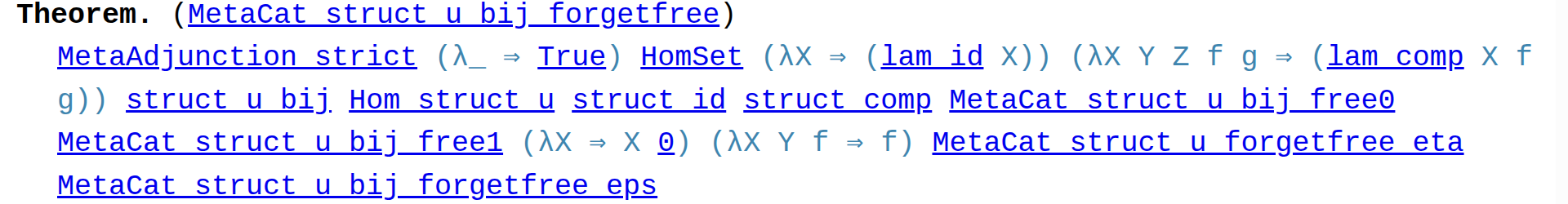}}
  \caption{\label{fig:bijfreethm} Adjunction Theorem for Bijections in the Megalodon Wiki}
\end{figure}

\begin{figure}[tb!]
  \centerline{\includegraphics[width=\textwidth]{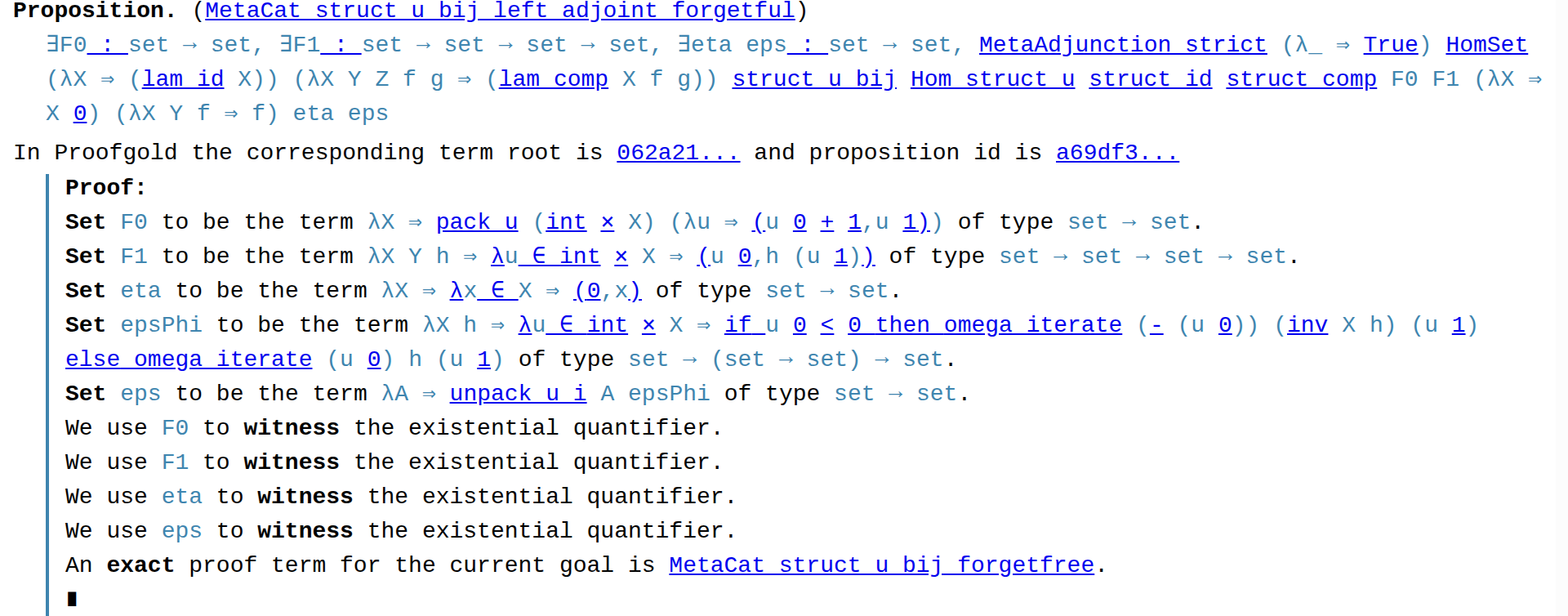}}
  \caption{\label{fig:bijfreeex} Existential Adjunction Theorem for Bijections in the Megalodon Wiki}
\end{figure}

The most recent of the 33 propositions to be proven has
$\cD$ as the category of structures with a carrier $X$
and a bijective function $f:X\to X$.\footnote{The proposition can be viewed on the explorer at the link
{\url{https://formalweb3.uibk.ac.at/pgbce/OP.php?b=a69df3cc99230330e94428aa4d4e3bf5ce0405944ff3242f3882144c1c0c5216}}.}
A proof of the corresponding proposition was published on the blockchain in June 2024.
The witnesses in the proof are as follows: 
\begin{itemize}
\item $\cF_0~X$ is the structure with carrier ${\mathbb{Z}}\times X$ (where ${\mathbb{Z}}$ is the set of integers)
  and the (bijective) unary function taking $u\in {\mathbb{Z}}\times X$ to $(u_0+1,u_1)$.
\item $\cF_1~X~Y~f$ is the morphism taking $u\in {\mathbb{Z}}\times X$ to $(u_0,f(u_1))\in {\mathbb{Z}}\times Y$.
\item $\eta~X$ is the function from $X$ to ${\mathbb{Z}}\times X$ taking $x$ to $(0,x)$.
\item $\varepsilon~(X,f)$ is the function from ${\mathbb{Z}}\times X$ to $X$ taking
  $u$ to $f^{u_0}(u_1)$. (Note that if $u_0 < 0$, this means iterating the inverse of
  the bijection $f$.)
\end{itemize}
There are definitions corresponding to these choices of $\cF_0$, $\eta$ and $\varepsilon$
in the Megalodon Wiki, and are shown here in Figure~\ref{fig:bijfreedefs}.
The main proof is of a theorem that checks these choices indeed give an adjunction.
The proof is long and detailed, so we only show the statement in Figure~\ref{fig:bijfreethm}.
The actual bounty was on the statement that such an adjunction exists.
Since all the details were checked in the previous theorem, we can show the
existential theorem with its proof in Figure~\ref{fig:bijfreeex}.

The most recent of the 33 propositions to be disproven takes $\cD$ to be
a category with monoids as objects.
Here ``disproven'' means the negation of the proposition was proven.
This should be surprising as it is clear that given a set $X$ one can
create a monoid freely generated by $X$, and this should provide the desired
left adjoint.
However, the category in question was defined to have semigroup homomorphisms
as arrows, and not every semigroup homomorphism (preserving the operation)
also preserves the identity element.
The definition is arguably a bug in the definition of the category of monoids.
The bounty presumably encouraged someone to look closely enough at the definition
to find and exploit the bug to prove the negation of the proposition.
Using the explorer to examine the proof of the surprising result,
others can discover the same bug.
The proof follows from the fact that there is no initial object in the category
of monoids and semigroup homomorphisms (and an alleged left adjoint would send the initial object 
in the category of sets to an initial monoid).
The proof that there is no initial monoid (with semigroup homomorphisms) essentially follows from the
fact that the two constant functions from a monoid to the
multiplicative monoid $\{0,1\}$ are both (always) semigroup homomorphisms, thus guaranteeing
an alleged initial monoid would not have a unique morphism to $\{0,1\}$. 

\begin{figure}[tb!]
  \centerline{\includegraphics[width=\textwidth]{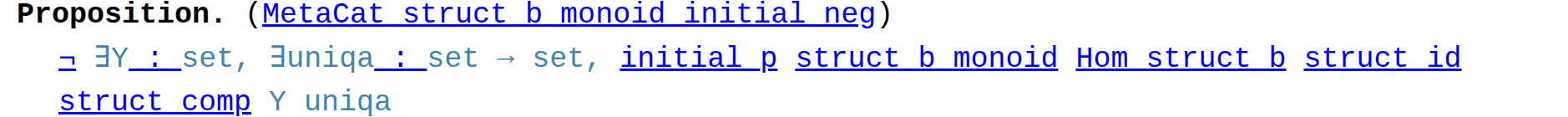}}
  \caption{\label{fig:noinitialmonoid} There is no Initial Monoid in the Megalodon Wiki}
\end{figure}

\begin{figure}[tb!]
  \centerline{\includegraphics[width=\textwidth]{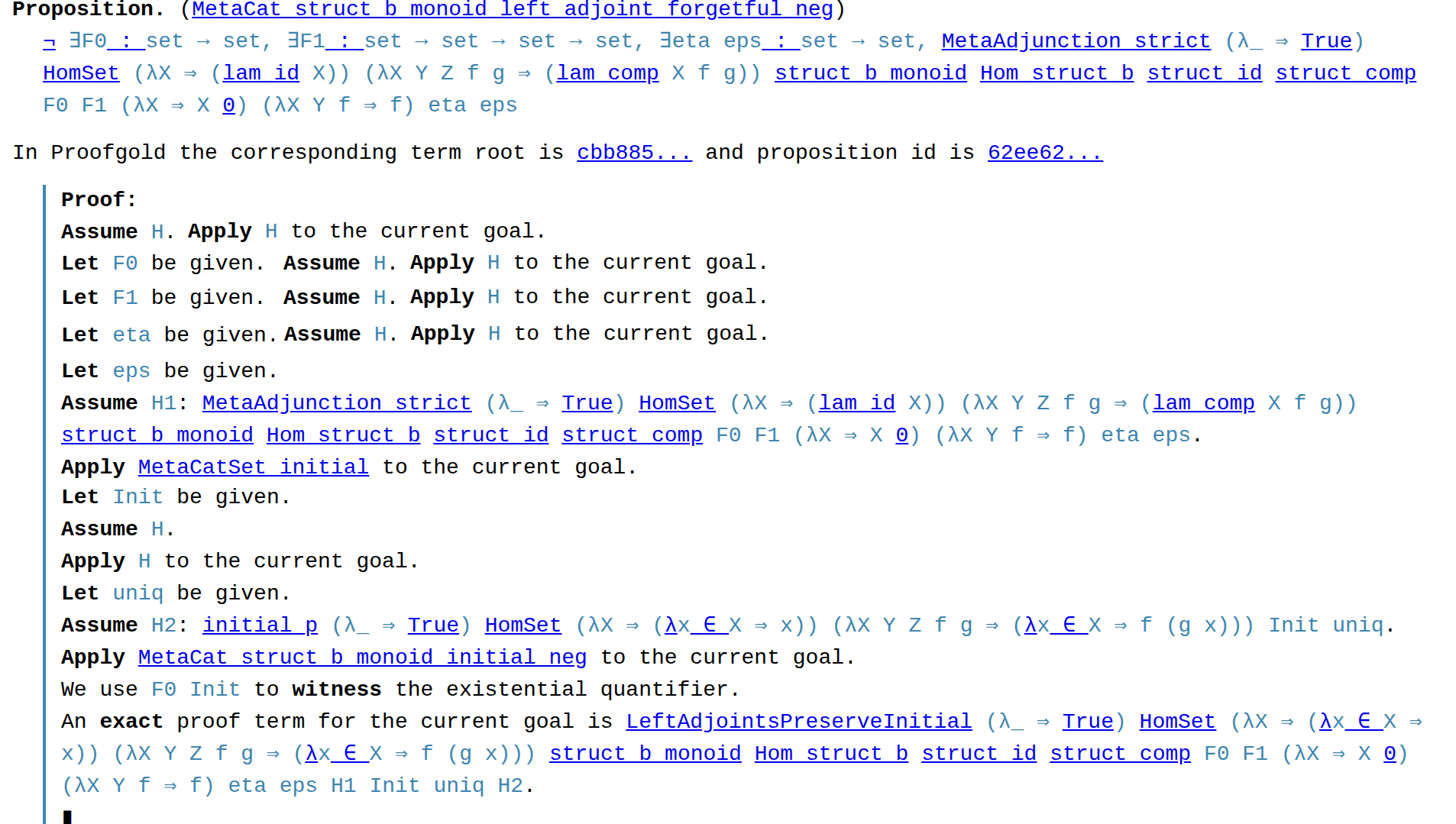}}
  \caption{\label{fig:noleftadjmonoid} There is no Left Adjoint to the Forgetful Functor for Monoids in the Megalodon Wiki}
\end{figure}

In a document in the Megalodon Wiki, the statement of the theorem that there is no initial monoid
(with semigroup homomorphisms)
is shown in Figure~\ref{fig:noinitialmonoid}. The proof (using the multiplicative monoid on $\{0,1\}$)
is also in the same document, but we omit it here as it is long and detailed.\footnote{Note that ${\mathsf{Hom\_struct\_b}}:\iota\invto\iota\invto\iota\invto o$
is defined so that ${\mathsf{Hom\_struct\_b}}~A~B~h$ holds
when $h$ is a function from the carrier of $A$ to the carrier of $B$ preserving the binary operation.
A correct definition of the category of monoids would extend the signature to explicitly include the
identity element in addition the binary operation. This would presumably have a name like ${\mathsf{Hom\_struct\_b\_e}}$.}
The negation of the adjunction conjecture is then proven as a consequence.
Its statement and proof in the document in the Megalodon Wiki are shown in
Figure~\ref{fig:noleftadjmonoid}.
The proof begins by assuming there exists an adjunction, i.e., $\exists F_0:\iota\invto\iota.\cdots$.
In general, a proposition $\exists x:\alpha.\varphi$ is considered the same as $\forall q:o.(\forall x:\alpha.\varphi\to q)\to q$.
This is why the existential assumption can be ``applied'' to the current goal (of proving $\bot$), followed
by a ``let'' (essentially giving a fresh name $x$ for the object) and an ``assume''
(giving the property $\varphi$ of the object $x$). This is repeated four times to obtain $F_0$, $F_1$, $\eta$
and $\varepsilon$. There is a previous (unshown) theorem that there is an initial object
in the category of sets. (It is witnessed by the empty set, but this is irrelevant here.)
We apply that previous theorem to obtain an initial set $Init$ and a function $uniq$ that
takes each set $X$ to the unique function from $Init$ to $X$.
Our goal is to prove $\bot$. We then apply the theorem from Figure~\ref{fig:noinitialmonoid}
which leaves the subgoal of proving there is an initial monoid.
The monoid $F_0~Init$ is used to witness that there is an initial monoid,
and this is justified by a previous (unshown) theorem that ensures left adjoints preserve initial objects.

An example of a proposition with an open bounty is given by taking $\cD$ to be
the category of groups.\footnote{The bounty can currently be seen on the explorer at the link
  {\url{https://formalweb3.uibk.ac.at/pgbce/q.php?b=TMQvwY1m9iU5rev4qXQjWWGYTZDHwCseEMv}}.}
In this case, the proposition should be provable
by taking $\cF_0~X$ to be the free group generated by $X$, but
no one has yet done this construction and proven the relevant theorem.


\section{Related Work}

Most proof assistants today are accompanied by tools for exploring their formal libraries via web interfaces.
The Mizar Mathematical Library (MML)~\cite{BancerekBGKMNP18}, one of the oldest central repositories of formalized mathematics,
in addition to its journal version, has been accessible through HTML renderings and search tools~\cite{FurushimaYSNW22,TSK23}.
Isabelle's Archive of Formal Proofs (AFP)~\cite{BlanchetteHMN15AFP} provides a curated collection of formal developments,
rendered online using Isabelle's document preparation system. The Coq proof assistant is supported by an online web
interface that allows interactive Coq sessions entirely within the web~\cite{CorbineauK07}. Its extension to ProofWeb
offers a web interface for multiple proof assistants, aiming to lower the barrier to entry for formal
verification~\cite{proofweb}. The Lean community has developed a variety of web tools, including the ProofWidgets library~\cite {NawrockiAE23},
with the paper discussing a lot of related work concerning user interfaces for theorem proving.

Beyond these assistant-specific interfaces, more advanced and general-purpose systems have been developed
to support exploration, integration, and sharing of formal mathematical knowledge. The \texttt{MathHub/MMT} framework
provides a logic-independent framework for querying across distributed libraries~\cite{BercicKR19}.
The Formal Abstracts aims to explore structured summaries of theorems that link informal and formal mathematics.
Its web access to both metadata and formal developments\footnote{\url{https://formalabstracts.github.io/}} has been impactful, but it appears to be currently inactive.

The Tezos~\cite{DoanT24} blockchain includes a smart contract programming language that has been designed to make
formal verification easier, which means that off-chain formal methods can be used to verify properties of programs/scripts.
This supports formal methods, albeit offline~\cite{BhargavanDFGGKK16}. Several blockchain explorers exist for Tezos, but they focus on the usual
smart contract information and do not include verification content.
Qeditas was an early attempt to combine formalized mathematics with blockchain technology,
but the project is no longer maintained; Proofgold builds on its ideas, refining and extending
them into a working system.

There have been a number of formalizations of category theory done in type-theoretic proof assistants; examples
include UniMath~\cite{AhrensMM19} in Coq, the Lean 3 library~\cite{Lean3}, Agda-Category~\cite{VezzosiM019};
several developments in Isabelle/HOL \cite{Stark16} and Isabelle/HOLZF and even Mizar~\cite{CAT_1.ABS}.
While type-theoretic systems allow concise and expressive formulations, they often provide less automation; in contrast, Isabelle/HOL offers strong automation but can make some constructions more cumbersome due to the lack of dependent types, and Mizar emphasizes human readability but it has limited expressiveness for higher-category theory.

\section{Conclusion}

We created a web-based blockchain explorer for Proofgold that allows users to interact with
mathematical knowledge contained there. We have demonstrated the utility of the system by
presenting several conjectures and their proofs or refutations using the system.
Future work includes improving support for faster collaboration, further improvements to the
visualization of the formal mathematics, as well as better search~\cite{AspertiGCTZ04}.

\paragraph{Acknowledgements} The results were supported by the Czech Ministry of Education, Youth and Sports within the dedicated program ERC CZ under the project POSTMAN no. LL1902, the ERC PoC grant \emph{FormalWeb3} no. 101156734, and the Czech Science Foundation grant no. 25-17929X.


\bibliographystyle{splncs04}
\bibliography{b}
\end{document}